\documentclass[a4paper]{article}
\usepackage{subfigure}
\usepackage{amssymb,amsthm,amscd, amsbsy, array}
 \usepackage{amsmath,bm,cite,color}
 \usepackage{yfonts}
 \usepackage{tikz}
   \usepackage{dsfont}
\usepackage{graphicx}
\newcommand{\mathsym}[1]{{}}

\usepackage{epsfig}
\usepackage{epsf}
\usepackage{rotating}
\usepackage{color}
\newfont{\tenmsb}{msbm10 scaled\magstep1}

\let\ssection=\section\renewcommand{\section}
{\setcounter{equation}{0}\ssection}

\newcommand{\mb}{\mathbf}

\newcommand{\lf}{\left (}

\newcommand{\rg}{\right )}

\def\smallover#1/#2{\hbox{$\textstyle{#1\over#2}$}}
\def\beq{\begin{equation}}
\def\eeq{\end{equation}}
\def\beq{\begin{equation}}
\def\eeq{\end{equation}}
\def\bea{\begin{eqnarray}}
\def\eea{\end{eqnarray}}

\def\lf{\left(}
\def\rg{\right)}
\def\lq{\left[}
\def\rq{\right]}
\def\lgr{\left\{}

\def\p{{\partial}}
\def\p{{\partial}}

\def\*{{\star}}

\def\xe{{{\'e }}}

\def\aa{{{\`a }}}

\def\chZ{{\u Z}}

\begin{document}
\title{Skyrmion States In Chiral Liquid Crystals}
\author{G. De Matteis $^{* \dag}$ ,$\quad$  L. Martina $^{ \dag \ddag}$, 
$\quad$ V. Turco $^\dag$
\\
 $^\dag$ Dipartimento di Matematica e Fisica, Universit\aa del Salento\\  
Via per Arnesano, C.P. 193 I-73100 Lecce, Italy\\
$^\ddag$ INFN, Sezione di  Lecce, Lecce, Italy  \\
$^*$ IISS "V. Lilla", MIUR,  Francavilla Fontana (BR)
\footnote{e-mail: giovanni.dematteis@unipv.it}
\footnote{e-mail:martina@le.infn.it}
 \footnote{ e-mail: vito.turco@live.com}}
 
\maketitle
\begin{abstract}
  Within the framework of Oseen-Frank theory, we analyse the static configurations  for chiral liquid crystals. In particular, we find numerical solutions for localised axisymmetric states in confined chiral liquid crystals with weak homeotropic anchoring at the boundaries. These solutions describe the distortions of two-dimensional skyrmions, known as either \textit{spherulites} or \textit{cholesteric bubbles}, which have been observed experimentally in these systems. Relations with nonlinear integrable equations have been outlined and are used to study asymptotic behaviors of the solutions. By using analytical methods, we build approximated solutions of the equilibrium equations and we analyse the generation and stabilization of these states in relation to the material parameters, the external fields and the anchoring boundary conditions. 
 \end{abstract}
 \section{Introduction}\label{sec:intro}
 
\begin{figure}
\centering
\subfigure[Cholesteric bubbles texture]{\includegraphics[width=7cm,height=5cm]{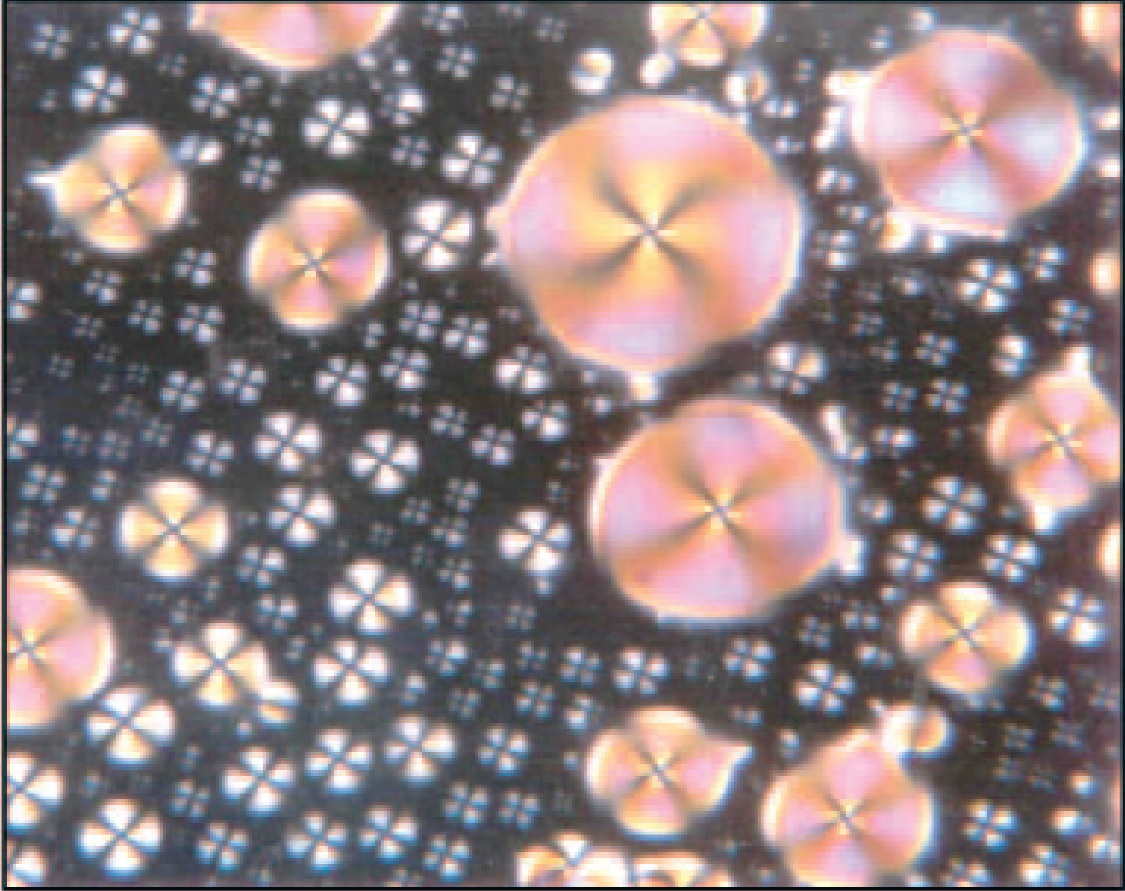}}
\subfigure[Cholesteric fingers texture]{\includegraphics[width=7cm,height=5cm]{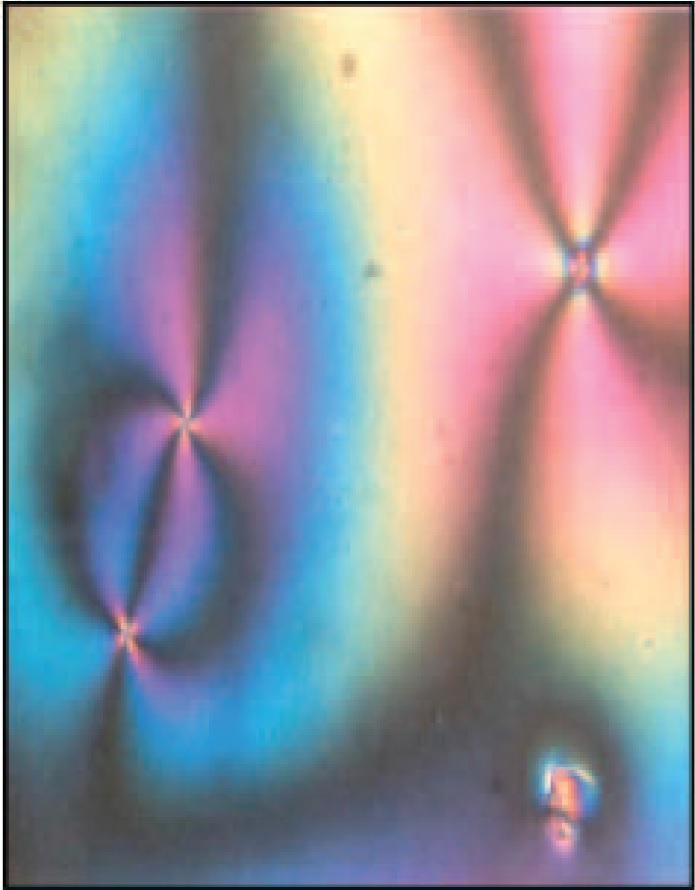}}
\caption{Pictures of the two textures observed in chiral nematics, obtained with polarized light microscopy\cite{carboni}.}
\label{fig:pics}
\end{figure}

 Recently a variety of 2-dimensional structures, \textit{cholesteric fingers}, and 3-dimensional ones, \textit{cholesteric bubbles} or \textit{spherulites}\cite{key-1}, have been observed in thin layers of Chiral Liquid Crystals (CLCs) with homeotropic anchoring on the confining surfaces. 
In particular, Carboni et al. detected a phase transition between the two textures, strongly depending on the thickness of the confining cell\cite{carboni}. They showed that the texture changes are  driven  by temperature through a parameter $\zeta$ proportional to the thickness and to a proper chirality parameter. Samples of different thickness displayed the textural changes at different
temperatures but for the same value of $\zeta$.
Pictures of the two phases, obtained with polarized optical microscopy, are shown in figure \ref{fig:pics}.
 In chiral systems of this kind, these isolated axisymmetric states are stabilized by specific interactions imposed by the underlying molecular handedness\cite{key-2}. Within the framework of the Frank-Oseen theory, we derive the equilibrium equations for these states and we study them through the use of numerical and analytical methods.
 
Free states cholesteric liquid crystals can be driven out of the equilibrium by applying external fields and by imposing anchoring boundary conditions \cite{1,2}. By experiencing both effects simultaneously, they are led to form  new structures, like cholesteric fingers \cite{3,8}, or helicoids, with defects disclination type and skyrmions \cite{9,12}, which are stabilized by topological and non-topological conservation laws and can be  described, at least in some approximate setting,  in terms of integrable nonlinear equations \cite{13,17}.

The present paper is organised as follows. After recalling the foundations of static continuum theory for 
chiral liquid crystals, in section \ref{sec:skyrme}, we will find and describe the skyrimion equilibrium configurations above mentioned  under the name of cholesteric bubbles. We will describe in detail the mechanisms which cause their generation and stabilization, for which the anchoring boundary conditions play a crucial role.\\
 Finally, in section \ref{sec:conclusion}, we sum up the obtained results and suggest a possible way to tackle the problem of finding analytical expressions of  helicoidal equilibrium configurations in the presence of an external electric field, and possible further developments for the presented analysis.
\section{Skyrmions in chiral liquid crystals}\label{sec:skyrme}
Let us consider a static cholesteric liquid crystal confined within the region $\mathcal{B}=\lbrace (x,y,z)\in\mathbb{R}^3,  \mid z\mid\leq \dfrac{L}{2}\rbrace$.\\
The system is described by a uni-modular director field $\mathbf{n}\lf \mathbf{r} \rg$ belonging to $\mathbb{RP}^2$\cite{deGennes,Stewart}, which   in polar representation has the expression 
\begin{equation}
\mathbf{n}(\mb{r})=(\sin\theta(\mathbf{r})\cos\psi(\mathbf{r}), \sin\theta(\mathbf{r})\sin\psi(\mathbf{r}), \cos\theta(\mathbf{r})), \qquad - \mathbf{n} \equiv \mathbf{n}. \label{directorpolar}
\end{equation}
In the bulk the liquid crystal director field $\mathbf{n}\lf \mathbf{r} \rg$ is governed by the Frank-Oseen  free energy $E_{FO}$ which reads as
\bea E_{FO}=\int_{\mathcal{B}} d^3 x \hspace{.2cm}\omega_{FO}[\mathbf{n}(\mathbf{x})]  \label{FrankOseenEn}
\eea
where
\bea
\omega_{FO}=    \frac{K_1}{2}(\nabla\cdot\mathbf{n})^2+\frac{K_2}{2}(\mathbf{n}\cdot\nabla\times\mathbf{n}-q_0)^2 +\frac{K_3}{2} (\mathbf{n}\times\nabla\times\mathbf{n})^2\nonumber\\
+\frac{(K_2+K_4)}{2}\nabla\cdot [(\mathbf{n}\cdot\nabla)\mathbf{n}-\nabla\cdot\mathbf{n}-(\nabla\cdot\mathbf{n})\mathbf{n}] -\frac{\varepsilon}{2}(\mathbf{n}\cdot\mathbf{E})^2 
. \label{fomega}
\eea
In \eqref{fomega} $q_0$ is the chirality parameter of the cholesteric phase, the positive reals $K_1$, $K_2$, $K_3$, $ K_4$ are the Frank elastic constants. The last term represents the interaction energy density associated with an external  static electric field $\mathbf{E}$, spatially uniform, along the $\mb{k}$ direction.  Of course, in the presence of the external electric field, the general rotational symmetry is broken and reduced to rotations around the direction of $\mathbf{E}$.  In the absence of anchoring conditions, the  field $\mathbf{n}\lf \mathbf{r} \rg$ would form a cholesteric helix with axis orthogonal to $\mathbf{E}$.  However, because of the bounding surfaces in the $\mathbf{k}$ direction,   the translational symmetry in the direction of $\mb{k}$ is broken, so that helices are deformed and confined within $\mathcal{B}$. Possibly extended structures called  helicoids, helicons (sometimes \emph{fingers}) and spherulites (also \emph{skyrmions}) can also form, depending on the existence of a preferred direction of perturbations of $\mb{n}$ in the directions orthogonal to $\mathbf{k}$. 

In order to calculate the structure and energy of such perturbations, we must minimise the Frank free energy under the appropriate boundary conditions. We also consider the simplifying  one constant approximation, i.e. we set
\beq  K = K_1 = K_2 = K_3, \hspace{.3cm}  K_4=0. \eeq
Corrispondingly, expression \eqref{FrankOseenEn} can be written as 
\beq E_{FO}=\int d^3 x \frac{K}{2}\left( \mid \nabla \mb{n}\mid^2-2 q_0 \mb{n}\cdot\nabla\times\mb{n} - \frac{\varepsilon}{K}(\mathbf{n}\cdot\mathbf{E})^2\right),\label{FrankOseenEn2}\eeq
where we used the well known identity 
\beq \nonumber (\nabla\cdot\mathbf{n})^2+(\mathbf{n}\cdot\nabla\times\mathbf{n})^2 + (\mathbf{n}\times\nabla\times\mathbf{n})^2 +\nabla\cdot [(\mathbf{n}\cdot\nabla)\mathbf{n}-\nabla\cdot\mathbf{n}-(\nabla\cdot\mathbf{n})\mathbf{n}] =\mid \nabla \mb{n}\mid^2.
\eeq
As far as the boundary conditions are concerned, we suppose that an homeotropic anchoring holds. Such a kind of conditions can be encoded into a variational formulation, considering the following additional surface energy contribution 
\beq
\omega_s=\frac{1}{2} K_s (1+\alpha(\mb{n}\cdot\bm{\nu})^2),
\eeq
where $K_s,\hspace{.1cm}\alpha>0$ and $\bm{\nu}$ being the unit outward normal to the boundary surface. Thus, the energy takes the expression $E_{FO}=\int d^3x\left(\omega_{FO}+\omega_s\right)$. Such additional term was first proposed by Rapini and Papoular\cite{rapini}. When $K_s\to\infty$ one can speak of strong homeotropic anchoring, which means the surface effects are taken into account in the form of Dirichlet boundary conditions 
\beq \mathbf{n}\lf x, y, z = \pm \frac{L}{2}\rg = \mathbf{k} \equiv  -  \mathbf{k},\label{surfcond2}\eeq
without any surface-related contribution in the expression for energy \eqref{FrankOseenEn2}.

In the following, we will describe the mechanisms, which give rise to skyrmionic and helicoidal perturbations when the liquid crystals are frustrated by the above confinement geometrical conditions.

We consider the director $\mb{n}$ in the form of equation \eqref{directorpolar}. Substituting such expression in \eqref{FrankOseenEn2},
 the Frank-Oseen free energy density functional will depend on the two scalar fields $\theta(\mb{x})$, $\psi(\mb{x})$ and their derivatives.
 
 We limit ourselves to axisymmetric isolated solutions, so we assume that $\theta=\theta(\rho,z)$ and $\psi=\psi(\phi)$, where $\rho$,$z$ and $\phi$ are the usual cylindrical coordinates around the axis $\mb{k}$. Thus, expression \eqref{FrankOseenEn2} becomes
 \beq
 \begin{split}
E_{FO}=&\frac{K}{2}\int_0^{2\pi}d\phi\int_{-\frac{L}{2}}^{\frac{L}{2}} dz\int_0^\infty\rho d\rho\Big[
\left(\frac{\partial \theta}{\partial z}\right)^2+\left(\frac{\partial \theta}{\partial \rho}\right)^2+\frac{\sin^2\theta}{\rho^2}\left( \frac{\partial \psi}{\partial \phi }\right)^2+\frac{\varepsilon E^2}{K}\sin^2\theta \\
+& 2q_0\left[\left(\frac{\partial \theta}{\partial \rho}\right)+\frac{\sin\theta\cos\theta}{\rho}\left(\frac{\partial\psi}{\partial\phi}\right)\right]\sin(\psi-\phi)+\omega_s(\theta)\Big]\label{polarFO},
\end{split}
 \eeq
where
\beq \omega_s(\theta)=\frac{K_s}{K}\sin^2\theta\delta\left( z\pm\dfrac{L}{2}\right)\label{surfcondsfer}\eeq
is the Rapini-Papoular energy contribution in the new system of coordinates.

The Euler-Lagrange equation associated with \eqref{polarFO} for $\psi$ is 
\beq
\label{eqpsi1}
2\frac{\sin^2\theta}{\rho^2}\psi_{\phi\phi}-2q_0\left[\theta_\rho+\frac{1}{2}\frac{\sin 2\theta}{\rho}\psi_\phi\right]\cos(\psi-\phi)=0.
\eeq
The solution to \eqref{eqpsi1} which minimises the energy \eqref{polarFO} is
\beq
\psi(\phi)=\phi+\frac{\pi}{2},\hspace{.5cm} \phi\in[0,2\pi]\label{eqpsi2}.
\eeq
Substitution \eqref{eqpsi2} into \eqref{polarFO} yields the Euler-Lagrange equation for the field $\theta(\rho,z)$ \cite{key-1}
\beq
\label{eqtheta2D}
\frac{\p^2 \theta}{\p z^2}+\frac{\p^2 \theta}{\p \rho^2}+\frac{1}{\rho}\frac{\p \theta}{\p \rho} -\frac{1}{\rho^2}\sin\theta\cos\theta
-\frac{2q_0}{\rho}\sin^2\theta-\frac{\varepsilon E^2}{K}\sin\theta\cos\theta=0.
\eeq
Since we are looking for finite energy solutions for $\theta\in [0,\pi]$, we impose the radial boundary conditions $\theta(\infty,z)=0$ and  $\theta(0,z)=\pi$. One can choose the alternative boundary conditions $\theta(\infty,z)=\pi$ and $\theta(0,z)=0$, in correspondence of the transformation $q_0\to-q_0$ in \eqref{polarFO}. Indeed, the sign of $q_0$ determines the handedness of the configuration $\theta$ which minimises the energy \eqref{polarFO}. From $\omega_s$ the boundary conditions at the planar confining surfaces are directly obtained  as
\beq
\label{surfcond}
\theta_z\left(\rho,\pm\frac{L}{2}\right)=\mp\frac{K_s}{2K}\sin 2\theta\left(\rho,\pm\frac{L}{2}\right).
\eeq
 We note that these conditions involve both $\theta$ and its derivative with respect to $z$.
 
It is convenient to rescale the equation and the boundary conditions with respect to the quantity $p=\frac{2\pi}{\mid q_0\mid}$, thus obtaining the adimensional BVP
 \begin{eqnarray}
\label{theta2Dscal}
\frac{\p^2 \theta}{\p z^2}+\frac{\p^2 \theta}{\p \rho^2}+\frac{1}{\rho}\frac{\p \theta}{\p \rho} -\frac{1}{\rho^2}\sin\theta\cos\theta
\mp \frac{4\pi}{\rho}\sin^2\theta-\pi^4\left(\frac{E}{E_0}\right)^2\sin\theta\cos\theta=0,
\vspace{1cm}\\
\begin{cases}
\label{bccases}
&\theta(0,z)= \pi,\vspace{.5cm}\hspace{.5cm}\theta(\infty,z)= 0,\\
&\p_z\theta\left(\rho,\pm\frac{\nu}{2}\right)=\mp 2\pi k_s \sin\theta\left(\rho,\pm \frac{\nu}{2}\right)\cos\theta\left(\rho,\pm \frac{\nu}{2}\right),
\end{cases}
 \end{eqnarray}
 where $E_0=\dfrac{\pi \mid q_0\mid}{2} \sqrt{\dfrac{K}{\varepsilon}}$ is the critical unwinding field for the  cholesteric-nematic transition in non-confined CLCs\cite{stewarta}, $\nu=L/p$ and $k_s=K_s/(K q_0)$; the $\pm$ sign in equation \eqref{theta2Dscal} depends on the sign of $q_0$. In the following we assume, with no loss of generality, $q_0<0$. 
 
 
 \subsection{Analytical analysis of Skyrmion solutions}\label{sec:analytical}
First of all, let us consider the radial reduction of \eqref{theta2Dscal}, i.e. $\theta_z=0$: 
\beq
\label{bulkscal}
\frac{\p^2 \theta}{\p \rho^2}+\frac{1}{\rho}\frac{\p \theta}{\p \rho} -\frac{1}{\rho^2}\sin\theta\cos\theta\mp \frac{4\pi}{\rho}\sin^2\theta-\pi^4\left(\frac{E}{E_0}\right)^2\sin\theta\cos\theta=0.
\eeq
with boundary conditions
\beq
\label{bordo21}
\theta(0)=\pi,\hspace{.5cm} \theta(\infty)=0.
\eeq
Equation \eqref{bulkscal} can not be solved analitically. However we can provide approximate analytical solutions. 
Note that if both $q_0\to 0$ and $E\to 0$ in equation \eqref{eqtheta2D}, equation \eqref{bulkscal} reduces to the Euler-Lagrange equation of the conformally invariant O(3)-sigma model in polar representation\cite{manton}, that is
\beq
\label{eqo3}
\frac{\p^2 \theta}{\p \rho^2}+\frac{1}{\rho}\frac{\p \theta}{\p \rho} -\frac{1}{\rho^2}\sin\theta\cos\theta=0.
\eeq 
Solutions of this model are well known and they date back to the work of Belavin-Polyakov\cite{belp}. They read as follows 
\beq
\label{1BP}
\theta=\arccos\left(\frac{\tilde{\rho}^2-4}{\tilde{\rho}^2+4}\right),
\eeq
with $\tilde{\rho}=\dfrac{\rho}{\rho_0}$ where $\rho_0$
is an arbitrary scale factor due to the conformal invariance.
\\
The fourth and the fifth term in \eqref{bulkscal} break the conformal symmetry, thus stabilising skyrmion solutions by lowering their energy and setting the scale factor $\rho_0$.
The two symmetry-breaking terms actually modify the Belavin-Polyakov vortex solution around $\rho=0$ and the behavior around $\rho\to\infty$, respectively. More specifically, the external electric field affects the shape of Skyrmion solutions as $\rho\to\infty$ since, in this limit, equation \eqref{bulkscal} reduces asymptotically to
\beq
\label{asint1}
\theta_{\rho\rho} - \pi^4\left(\frac{E}{E_0}\right)^2\sin\theta\cos\theta=0.
\eeq
The resulting asymptotic behavior is 
\beq
\label{asint2}
\theta(\rho)\leadsto e^{-\frac{\rho}{\rho_1}},\qquad\text{as}\quad \rho\leadsto\infty,\qquad \text{with} \hspace{.2cm} \rho_1=\frac{1}{\pi^2}\frac{E_0}{E},
\eeq
which shows that $\theta$ is exponentially decaying to zero in this limit.

In order to explore the behavior of the solution in a larger neighborhood of $\rho\leadsto\infty$ as well as $\theta$ around zero, we can now proceed with the linear approximation of equation \eqref{bulkscal} which, at the first order in $\theta$, yields to
the modified Bessel equation\cite{gradshtein}
\beq
\rho^2\frac{\p^2 \theta}{\p \rho^2}+\rho\frac{\p \theta}{\p \rho}-(1+\pi^4\left(\frac{E}{E_0}\right)^2\rho^2) \theta=0.
\eeq
Its general solution is
\beq
\label{lin1D}
\theta(\rho)=c_1 I_1\left(\frac{\rho}{\rho_1}\right)+c_2 K_1\left(\frac{\rho}{\rho_1}\right),
\eeq
where $I_1$ e $K_1$ are known as the first order modified Bessel functions of first and second kind, respectively, with $c_1$ e $c_2$ arbitrary constants depending on boundary conditions.\\
Function $K_1$ has the correct asymptotic behavior as $\rho\rightarrow \infty$, but it diverges at the origin. On the other hand, $I_1\left(\dfrac{\rho}{\rho_1}\right)\xrightarrow[\rho \to 0]{} 0$ but $I_1\left(\dfrac{\rho}{\rho_1}\right)\xrightarrow[\rho \to +\infty]{} +\infty$, so that funtion $I_1$ cannot approximate the solution we are looking for.

It is now clear that \eqref{lin1D} cannot be an approximate solution to \eqref{bulkscal} for all $\rho$, except for $\rho\to\infty$ where \cite{gradshtein} (eq. 8.456)
\beq
\label{asymptlin}
\theta\leadsto c_2\sqrt{\dfrac{\rho_1}{\rho}} \exp\left[-\dfrac{\rho}{\rho_1}\right].
\eeq
 If we now consider as dominant the interaction with the external electric field with respect to the chiral term, we obtain from equation \eqref{bulkscal} a new non-linear asymptotic approximation
\beq
\label{sgcil}
\frac{\p^2 \theta}{\p\rho^2}+\frac{1}{\rho}\frac{\p\theta}{\p \rho}-\frac{\rho_1^2}{2} \sin 2\theta=0,
\eeq
which is known as cylindrical Sine-Gordon equation \cite{barone}. The most relevant fact about it is the connection with the celebrated Painlev\xe equations\cite{ablowitz,clarkson}. In particular, cylindrical Sine-Gordon equation was first connected to the Painlev\xe $III$ in the work \cite{McCoy} by applying the transformation 
\beq
\theta(\rho)=-i \hspace{.15cm}\text{ln}\left( \frac{q(t)}{\sqrt{t}}\right), \hspace{.5cm} t=\left(\frac{\rho}{\rho_1}\right)^2
\eeq
and obtaining
\beq
q^{\prime\prime}=\frac{1}{q} q'^2-\frac{1}{t}q'+\frac{q^3}{16 t^2}-\frac{1}{16 q},
\eeq
which is a particular case of the general Painlev\xe $III$
\beq
\label{P3eq}
q^{\prime\prime}=\frac{1}{q} q'^2-\frac{1}{t}q'+\frac{q^2 (a+c q)}{4 t^2}+\frac{b}{4 t} + \frac{d}{4 q},
\eeq
where $a$,$b$,$c$,$d$, are complex arbitrary constants.
Equation \eqref{P3eq} was first integrated in \cite{nakamura} and the asymptotics of the solutions of equation \eqref{sgcil} were analysed in \cite{fokasits}. It has the general solution parametrised by two complex Cauchy data, say $\alpha, \beta$, in such a way that $\theta\lf \rho | \alpha, \beta\rg$  has the asymptotic behavior
\beq\label{asymptzero}
\rho\leadsto 0 \text{:}\hspace{0.5cm} \theta(\rho)\leadsto \alpha \hspace{.1cm}\text{ln}\left(\frac{\rho}{\rho_1}\right)+i\frac{\pi}{2}\alpha+\beta + O\left( \left(\frac{\rho}{\rho_1}\right)^{2-\mid \Im\alpha\mid}\right), \qquad  (|\Im\alpha| < 2)
\eeq
and
\beq\label{asymptSGcil}\rho \leadsto \infty \qquad \theta  \left( \rho \right) \leadsto
   \left[ b_+
   e^{\frac{\rho }{\rho _1}}
   \left(\frac{\rho }{\rho_1}\right){}^{-\frac{1}{2} +  i \omega} + b_-
   e^{-\frac{\rho }{\rho _1}}
   \left(\frac{\rho }{\rho
   _1}\right){}^{-\frac{1}{2} -
   i \omega}\right] \left(O\left(\frac{\rho _1}{\rho
   }\right)+1\right) + O\left(   \left(\frac{\rho }{\rho_1}\right){}^{3
   |\Im \omega|-\frac{3}{2}}\right),\eeq
where the constant $b_{\pm}$ and $\omega$ are related to the Cauchy data by  the {\it connection formulas} determined in \cite{Novokshenov}
\beq  e^{-\pi  \omega }  \sin (2 \pi  \sigma
   ) = \sin (2 \pi 
   \eta )
\eeq   
and
\beq \label{b+b-} b_+ =
   \frac{-2^{2 i \omega
   } e^{-\pi  \omega }}{\sqrt{\pi
   }} \Gamma (1-i \omega ) \frac{\sin
   (2 \pi  (\eta +\sigma
   ))}{\sin (2 \pi  \eta
   ) },\quad 
   b_- = 
   \frac{i \, 2^{-2 i \omega} }{\sqrt{\pi
   }}
   \Gamma
   (1 + i \omega ) \frac{\sin
   (2 \pi  (\eta - \sigma
   ))}{\sin (2 \pi  \eta
   ) }, \eeq
   with
   \beq\label{eqsigma}\sigma = \frac{1}{4} + \frac{i}{8} \alpha , \quad \eta = \frac{1}{4} + \frac{1}{4\pi} \left( \beta + \alpha \, \ln 8\right) + \frac{i}{2 \pi} \ln \frac{\Gamma\left( \frac{1}{2} -\frac{i\, \alpha}{4}\right)}{\Gamma\left( \frac{1}{2} + \frac{i\, \alpha}{4}\right)}.
   \eeq
 These constants satisfy the relation 
   \beq\label{omegacil} b_-  b_+ = - 4 i \omega, \qquad |\Im \omega | < \frac{1}{2}. \eeq
   
   From the exponential decaying of $\theta$ at $\rho \to \infty$ obtained in the linear approximation, one has to fix $b_+ = 0$.  Then, $\omega = 0 $ because of \eqref{omegacil}. Taking into account the other relations one gets to the  relation
   \beq\label{eqeta}\eta = - \sigma + \frac{1}{2} + k, \quad k \in \mathbb{Z}.\eeq
   This leads to set $b_- = -2 i \sqrt{\frac{1}{\pi }}
   \cos (2 \pi  \sigma )$ and a relation is established between $\alpha $ and $\beta$, i.e.
   \beq \label{eqbeta}  \beta = -\left(  \frac{i \pi }{2} +\ln 8\right) \alpha  -2 i  \ln \frac{\Gamma\left( \frac{1}{2} -\frac{i\, \alpha}{4}\right)}{\Gamma\left( \frac{1}{2} + \frac{i\, \alpha}{4}\right)} +  4 k\pi . \eeq
   
   It is worth to specify that from equations \eqref{asymptlin}, \eqref{b+b-}, \eqref{eqsigma}, \eqref{eqeta} and \eqref{eqbeta} we can set the value of $\alpha$ to be
\beq
\alpha=- \frac{4}{\pi} \text{arcsinh} \left(\frac{\sqrt{\pi}}{2} c_2\right)\in\mathbb{R}^-,
\eeq
so that function $\theta$ as in (\ref{asymptzero})   takes his value in $\mathbb{R}$. The irregular behavior, i.e.  logarithmic divergence at $\rho\leadsto 0$, is a consequence of the approximation we make use of when we obtain equation \eqref{sgcil}, neglecting the chiral term. 



Since we fail in finding regular approximate solutions with standard methods, we turn our attention on the scaling-variational ansatz in \cite{key-1,sferuliti}. Exploiting the results obtained above, we examine this ansatz and we use it to build an approximate solution of \eqref{theta2Dscal}.  We then study the competitive effects of homeotropic anchoring and of the external eletric field on it.

Let us consider equation \eqref{bulkscal}, whose solutions $\theta(\rho)$ exhibit an exponential decay for large distances and a trend approximately linear for small distances. As we said, the behavior around $\rho\approx 0$ is sufficiently well described by the 1-vortex of Belavin-Polyakov. Substituting solution \eqref{1BP} in equation \eqref{bulkscal} we obtain the following condition
\beq
\label{328}
-\pi^3\left(\frac{E}{E_0}\right) ^2   \left(\rho^2-4
   \rho_0^2\right)\mp 16\rho_0
   =0.
\eeq
Around $\rho\to 0$, \eqref{328} leads to an extimation of $\rho_0$

\beq
\label{rho0}
\rho_0= \frac{4}{\pi^3} \left(\frac{E_0}{E}\right)^2=4\pi \rho_1^2,
\eeq
which can be interpreted as the typical scale of a spherulite. Then, around $\rho= 0$, the solution of \eqref{bulkscal} is approximated by the Belavin-Polyakov $1-$vortex with $\rho_0$ fixed by \eqref{rho0}, which at first order in $\rho$ takes the form
\beq
\label{ansatzbulk}
\theta (\rho)= \pi-\frac{\rho}{\rho_0}+O\left(  \left(\frac{\rho}{\rho_0}\right)^3\right).
\eeq
\begin{figure}[h]
\centering
\includegraphics[width=10cm, height=4cm]{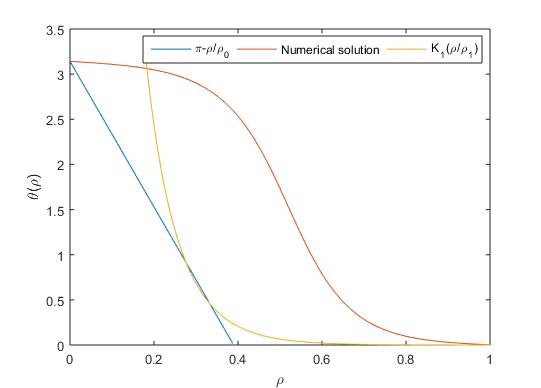}
\caption{Comparison between the numerical solution of \eqref{bulkscal} and the analytcal linear approximations for $\frac{E}{E_0}=1.02$.}
\label{fig:comparison1.02Ana}
\end{figure}
\begin{figure}[h]
\centering
\includegraphics[width=10cm, height=4cm]{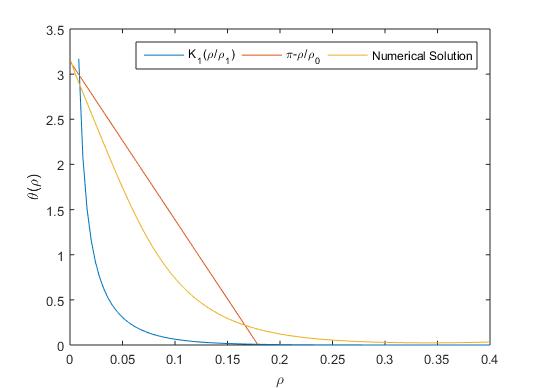}
\caption{Comparison between the numerical solution of \eqref{bulkscal} and the analytcal linear approximations for $\frac{E}{E_0}=1.5$}
\label{fig:comparison1.5Ana}
\end{figure}
 For sufficiently large  electric fields, i.e. $\frac{E}{E_0} > 1$, around $\rho = 0$  and $\rho \to \infty$ the linear approximations  match with the numerical solution quite closely, as represented in fig. \ref{fig:comparison1.02Ana}. On the other hand, the approximations become very rough for relatively weak fields , i.e. $\frac{E}{E_0}  \approx 1$, as shown in fig. \ref{fig:comparison1.5Ana}. As far as  the numerical cases considered in the present work, this behaviour denotes the underestimation of the  chiral term in the linear approximation, in particular in the intermediate scales $\rho_1\leq \rho  \leq \rho_0$.  Furthermore, the use of the nonlinear approximation (at least for large electric fields) (\ref{sgcil})  does not seems to be  very helpful. In fact the $\log$ in (\ref{asymptzero}) pushes the region in which  $\theta$ take values near $\pi$ closer to 0 than the Bessel $K_1$ type solution (\ref{lin1D}) does. 

Now, looking for the $z$ dependence of the spherulites, we adapt to our analysis the method suggested in  \cite{key-1}, supposing  that the more relevant contribution to  the free energy  of equation \eqref{theta2Dscal}  comes from a neighbourhood of $\rho = 0$.  There a solution  $\theta(\rho, z)$ of  \eqref{theta2Dscal}  is guessed  \cite{sferuliti}  to be weakly modulated by a  $z$-scaled dependence  on $\rho$ in the form 
\beq
\label{scalz}
\theta (\rho , z)=\pi-\tilde{\theta}\left( \frac{\rho}{Z(z)}\right)
\eeq
for a suitable $\tilde{\theta}$.  Accordingly to  \eqref{ansatzbulk}, a linear approximation   the expression (\ref{scalz}) has to be
\beq \label{linansatz}
\theta (\rho ,z)= \lgr 
\begin{array}{cc}
 \pi- \frac{\rho}{\rho_0 Z(z)} & 0 <  \rho/Z(z)< \pi\rho_0   , \\
  0 &     \rho/Z(z)>\pi\rho_0  
\end{array}\right.
\eeq
with $\rho_0$ given \eqref{rho0}. 

By using equation \eqref{scalz}, in units of $K$  the free energy \eqref{FrankOseenEn2} can be rewritten as
\beq
\overline{E}=I_0\int_{-\nu/2}^{\nu/2}dz \left[ \left( \frac{d Z}{dz}\right)^2 + A\pi^4 \left(\frac{E}{E_0}\right)^2 Z^2- B 4\pi Z + k_s A Z^2\delta(z\pm \frac{L}{2})\right],
\eeq
where 
\beq
A= \frac{I_1}{I_0},\;B= \frac{I_2}{I_0},\; \textrm{with}\;
\label{coefficientiint}
I_0=\int_0^\infty \left( \frac{d \theta}{d \rho}\right)^2\rho^3d\rho, \hspace{.3cm} I_1=\int_0^
\infty \sin^2\theta\rho d\rho, \hspace{.3cm} I_2=\int_0^\infty \left( \frac{d\theta}{d\rho}+\frac{\sin\theta\cos\theta}{\rho}\right)\rho d\rho ,
\eeq
and where the conformal invariance of $(\nabla \theta)^2$ has been taken into account.

\begin{figure}[!ht]
\centering
\subfigure[Contour plot of solution \eqref{linansatz} for $(E/E_0, \nu,k_s)=(1.02, 1.8, 0)$.]
{\includegraphics[width=5cm, height=4cm]{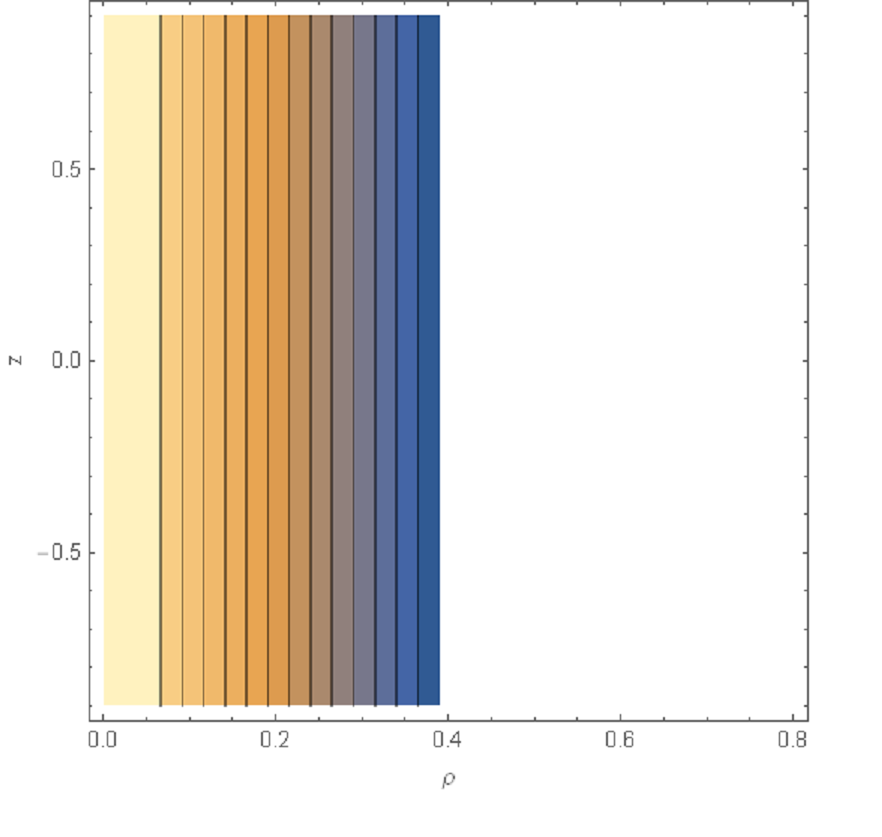}}
\subfigure[Contour plot of solution \eqref{linansatz} for $(E/E_0, \nu,k_s)=(1.02, 1.8, 1.5)$.]
{\includegraphics[width=5cm, height=4cm]{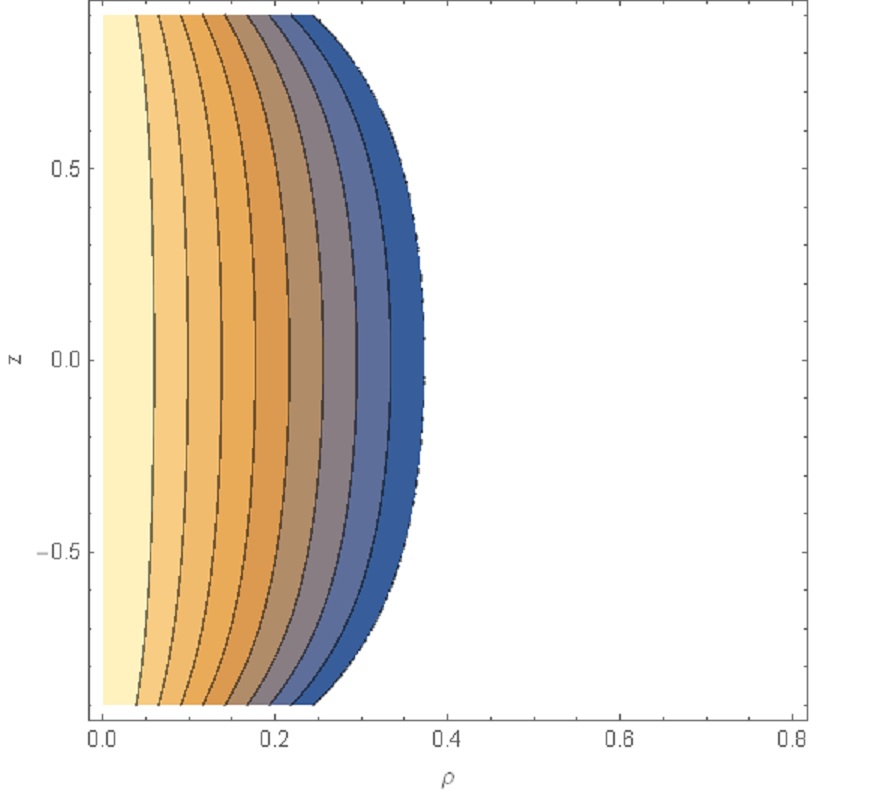}}
\subfigure[Contour plot of solution \eqref{linansatz} for $(E/E_0, \nu,k_s)=(1.02, 1.8, 5)$.]
{\includegraphics[width=5cm, height=4cm]{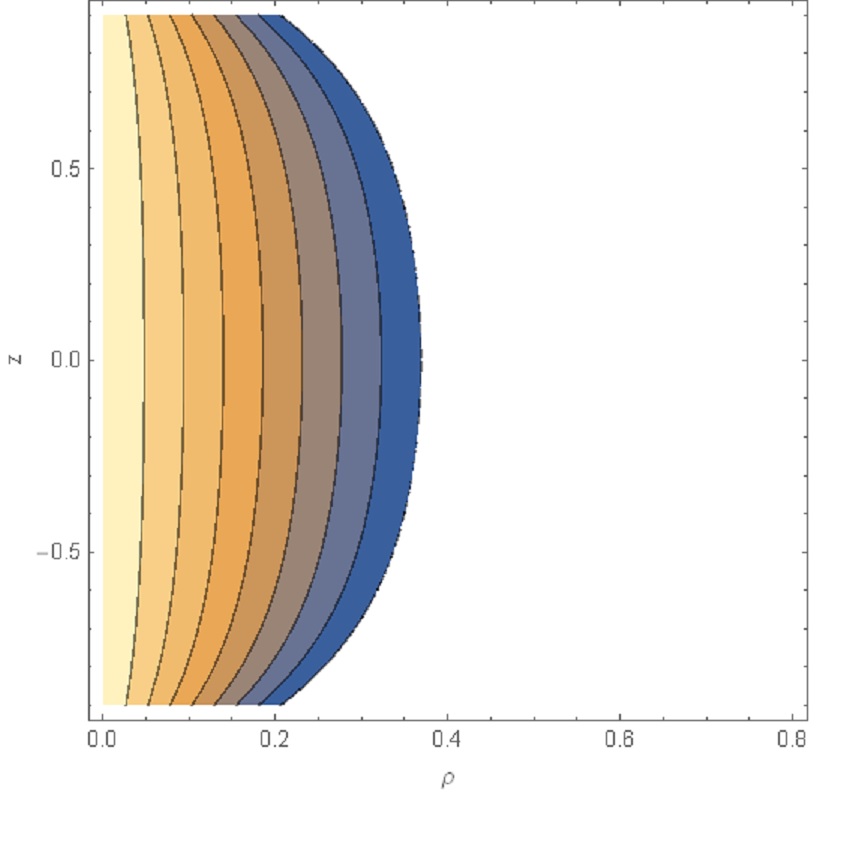}}
\caption{\textit{Contour plot of solution \eqref{linansatz} for three different values of $k_s$ with constant external electric field and cell thickness.}}
\label{fig:approxsfe}
\end{figure}

Using the expression \eqref{linansatz}, the integrals in \eqref{coefficientiint} can be evaluated explicitly in the interval $[0,\pi\rho_0 Z(z)]$ leading to the expression 
\beq
\label{overE1}
\overline{E}=\frac{\pi^4 }{4 }\rho_1^2 Z'(z)^2+\frac{1}{4} \pi ^2
   Z(z)^2-\frac{1}{2} \pi ^2 Z(z).
\eeq
From the expression \eqref{overE1} one can derive the Euler-Lagrange equation for the unknown $Z$, namely    
\beq
\label{eleqz}
 Z''(z)- \frac{1}{\pi^2  \rho _1^2} Z(z)+\frac{1}{\pi^2  \rho _1^2}=0,
\eeq
which has the general solution 
\beq
Z(z)=q_1 e^{-\dfrac{z}{\pi  \rho _1}}+q_2
   e^{\dfrac{z}{\pi  \rho _1}}+1, \quad q_i \in \mathbb{R}.
\eeq
Imposing the boundary conditions \eqref{surfcond},
 we obtain the approximate scaling factor
\beq
\label{zansatz}
Z(z)= 1-\frac{2 \pi k_s  \cosh \left(\frac{ z}{\rho_1}\right)}{2 \pi  k_s  \cosh \left(\frac{  \nu }{2 \rho_1
   }\right)+\frac{1}{\rho_1} \sinh \left({\frac{  \nu }{2 \rho_1 }}\right)}.
\eeq
We notice that the vortex size decreases as $\mid z\mid$ and $k_s$ increase, as it can be seen in figure  \ref{fig:approxsfe}. 

The availability of an analytic solution, even if approximated, allows us to study the energetic of the skyrmion states. In fact, considering the total energy \eqref{FrankOseenEn2} and substituting for $\theta(\rho,z)$ its expressions \eqref{linansatz} and \eqref{zansatz}, we obtain the estimation 
\beq
\begin{split}
\mathcal{E}_c(E,\nu,k_s)=&\pi  \biggl(-\text{Ci}(2 \pi ) \nu +\frac{32 \pi ^9\rho_1^3 k_s ^2 \sinh
   \left(\frac{\rho_1}{\pi}  \nu \right)}{ \left(2  k_s  \cosh \left(\frac{\rho_1}{\pi} \frac{ \nu }{2 
   }\right)+\frac{1}{\pi^2\rho_1}  \sinh \left(\frac{\rho_1}{\pi} \frac{\nu }{2 
   }\right)\right)^2}\\
   &-\frac{128 \pi ^9\rho_1^3 k_s }{ \left(2
    k_s  \coth \left(\frac{\rho_1}{\pi} \frac{  \nu }{2 
   }\right)+\frac{1}{\pi^2\rho_1} \right)}+24 \pi ^6\rho_1^2 \nu +\pi
   ^2 \nu +\gamma  \nu +\nu  \log (2 \pi )\biggr),
   \end{split}
\eeq
where $\text{Ci}$ is the  cosine integral function and $\gamma$ is the Eulero's constant.  \\
With a similar procedure,  for the  solution \emph{in the bulck} \eqref{ansatzbulk} we obtain for the energy  the expression 
\beq
\mathcal{E}_b(E,\nu)=\pi  \nu  \left(-\text{Ci}(2 \pi )+24 \pi ^6 \rho_1^2+\pi
   ^2+\gamma +\log (2 \pi )\right).
\eeq
 It's clear that, with fixed $E/E_0$ and $\nu$, it results $\mathcal{E}_c<\mathcal{E}_b$. In particular, as $k_s$ increases the gaping between the two energies grows up, while it tends to decrease as $E/E_0$ increases, showing a tendency of the LC to take the uniform ordering for high values of the electric field, against the distortion bue to the  chiral  and anchoring effects.

 \subsection{Numerical analysis of Skyrmion solutions}
 The Boundary Value Problem (BVP) \eqref{theta2Dscal} can be solved numerically, through the use of standard central finite difference discretization and the Newton-Raphson method. The problem can be coded in almost any programming language\cite{recipes, leveque}. However, we used  MATLAB\textregistered\hspace{.1cm} by Mathworks\cite{matlab} because it easily operates with large and sparse matrices. \\To find a suitable initial guess for the iterative method, we make use of a shooting method for the planar reduction of equation \eqref{eqtheta2D} (i.e. $\theta_z=0$) and we extend the resulting planar profile over the whole cell. \\
  
\begin{figure}[!h]
\centering
\subfigure[$\left(\dfrac{E}{E_0}\right) =1.02,k_s=0.1$]
{\includegraphics[width=.48\linewidth, height=5cm]{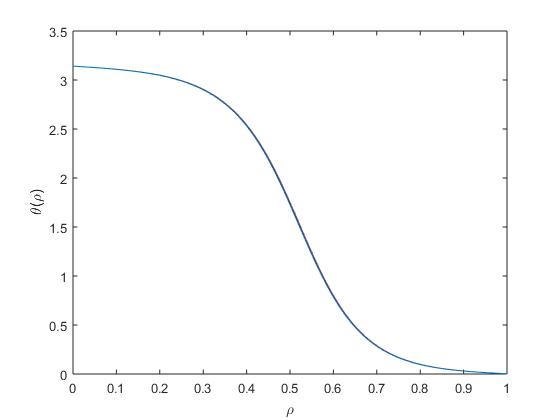}}
\subfigure[$\left(\dfrac{E}{E_0}\right) =1.02, k_s=6$]
{\includegraphics[width=.48\linewidth, height=5cm]{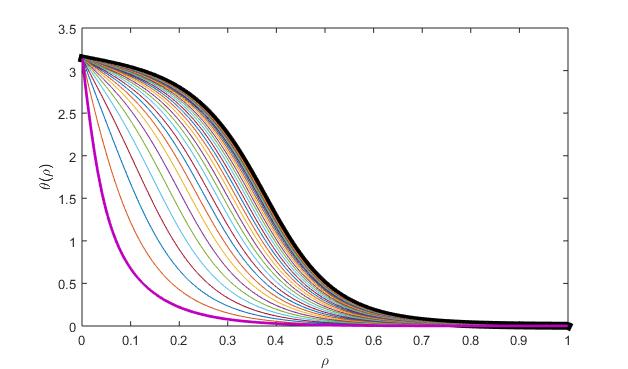}}
\caption{Profiles $\theta (\rho)$ for $E/E_0=1.02$. Different curves refer to different values of $\mid z \mid$.  Bold curves have to be referred  to $\mid z\mid=0$ (the black one) and to $\mid z	\mid=\nu/2$ (the purple one).}
\label{fig:profiles1}
\end{figure}

\begin{figure}[!h]
\centering
\subfigure[$\left(\dfrac{E}{E_0}\right) =1.5,k_s=0.1$]
{\includegraphics[width=.48\linewidth, height=5cm]{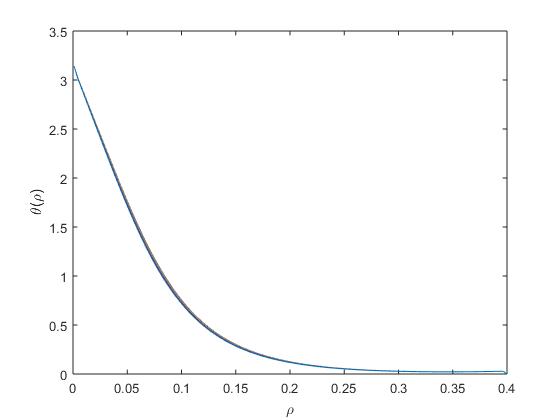}}
\subfigure[$\left(\dfrac{E}{E_0}\right) =1.5, k_s=6$]
{\includegraphics[width=.48\linewidth, height=5cm]{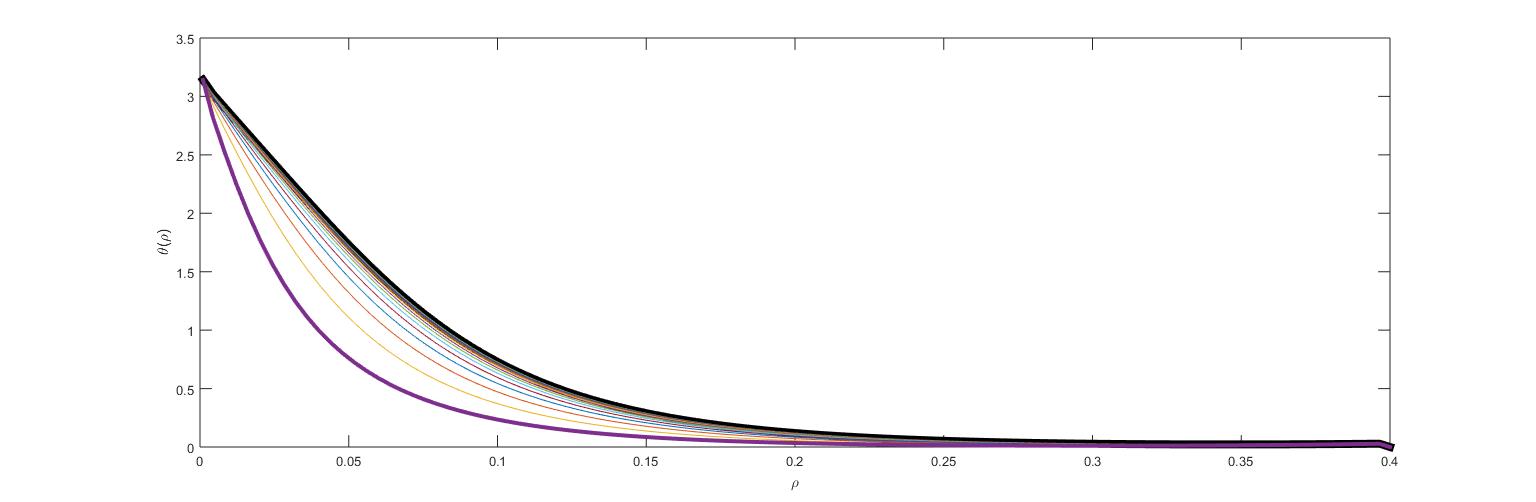}}
\caption{Profiles $\theta (\rho)$ for $E/E_0=1.5$. Different curves refer to different values of $\mid z\mid$.  Bold curves have to be referred  to $\mid z\mid=0$ (the black one) and to $\mid z\mid=\nu/2$ (the purple one). We note that the effect of a greater external electric field is to make smaller the size of the vortices, for fixed values of $k_s$.}
\label{fig:profiles2}
\end{figure}

 The numerical solutions of the BVP \eqref{theta2Dscal}, for different values of the couple $\left( \dfrac{E}{E_0}, k_s\right)$ are depicted in figures \ref{fig:profiles1} and \ref{fig:profiles2}. In each figure the profiles $\theta(\rho)$ for different values of $z\in [-\nu/2,\nu/2]$ are represented. In figure \ref{fig:profiles1} we have $\dfrac{E}{E_0}=1.02$ and the strength of the anchoring  $k_s=0.1, 6$. In figure \ref{fig:profiles2} we have $\dfrac{E}{E_0}=1.5$ with the same values for $k_s$. We note that, when the strength of the anchoring is small, the profiles are almost equal for any value of the coordinate $z$. This means that, when the interfaces at the boundaries of the cell have a really small homeotropic effect on the director's configuration, a quasi-perfect cylindrical symmetry holds for axisymmetric solutions. In this case, the planar vortices described by $\theta(\rho)$ for every value of $z$, have the same, maximum, size. However, if we impose a quite strong homeotropic effect at the boundaries, the vortices tend to have a reduced size, which becomes smaller as $\mid z\mid$ reaches the value $\dfrac{\nu}{2}$. In both figures \ref{fig:profiles1} and \ref{fig:profiles2}, the value of the dimensionless  thickness of the cell is $\nu=1.8$.
 
\begin{figure}
\centering
\subfigure[$\left(\dfrac{E}{E_0}\right) =1.02$]{\includegraphics[width=.6\linewidth,height=5.2cm]{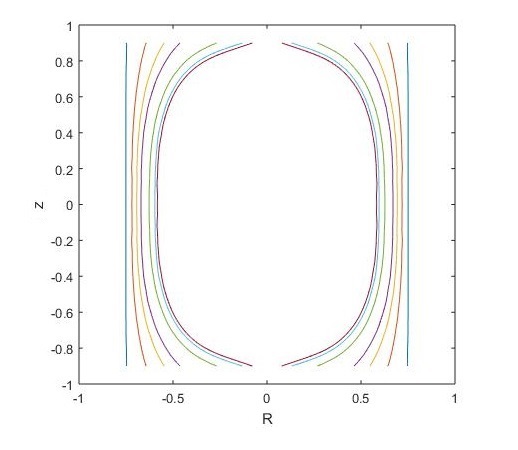}}
\subfigure[$\left(\dfrac{E}{E_0}\right) =1.5$]{\includegraphics[width=.6\linewidth,height=4.8cm]{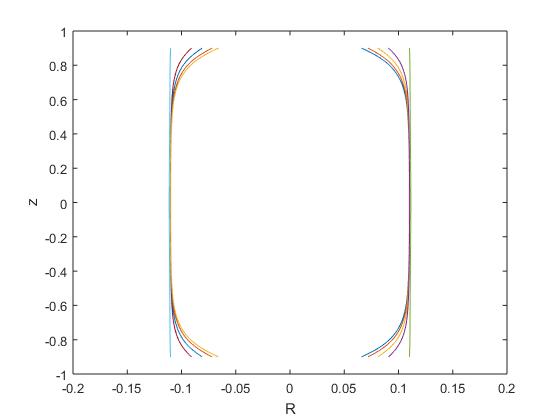}}
\caption{Size of the planar vortices for different values of $\mid z\mid$ . Different colors refer to different values of $k_s$ ($k_s=0.1,0.5,1,1.5,3,6,12$). We note that, as mentioned above, for $k=0.1$ a quasi-cylindrical symmetry holds for solutions, independently of the strength of the anchoring $k_s$. However, the structures tend to have the form of a bubble as $k_s$ increases.}
\label{fig:size1}
\end{figure}

Leonov et al. proposed a method to estimate  the size of the solutions of the BVP \eqref{theta2Dscal}, similarly   to what already done  for  the size of the ferromagnetic domain walls \cite{key-1,magdom}. Such a procedure consist of tracing the tangent at the inflection point $\rho_I\lf z \rg$ of $\theta(\rho, z)$, i.e. where $\theta_{\rho \, \rho}(\rho_I, z) = 0$  (numerically evaluated) for a fixed value of $z$. The point $R\lf z \rg$  in which that tangent intersects the $\rho$-axis  gives estimation, namely 
\beq
R\lf z \rg = \rho_I\lf z \rg-\theta (\rho_I, z) \theta_\rho (\rho_I, z )^{-1}.
\eeq   
The results of this procedure are reported in figures \ref{fig:size1}.a and \ref{fig:size1}.b, for the two different values of $\dfrac{E}{E_0}$ taken into consideration. We stress that for greater external fields  all vortices shrink .

\section{Conclusions}\label{sec:conclusion}
In the present work we gave a description of the isolated axisymmetric skyrmion states arising in confined chiral liquid crystals. We show how the interplay between chirality, external fields and homeotropic anchoring is responsible of their generation and stabilization.  The equilibrium equations for these states depend on the three material parameters: $E/E_0$, $k_s$, $\nu$, in the adimensional representation of the model introduced above. 
\\
We provided an analytical study of the linearly approximated equation \eqref{theta2Dscal}.   An interesting feature of such an equation is the link, even if in an approximate setting for large electric fields, with the Painlev\xe III examined in section  \ref{sec:analytical}. However, we showed that both linear and nonlinear, in the previous sense, approximations do not take into account the chiral effect, which is dominant at intermediate distances  for any values of the electric field, at least  below a  critical value leading to the uniform distribution of the order parameter. Thus, how to deal with such a nonlinear  interaction  by using analytical tools  seems to be an interesting challenge for future studies. 
\\
On the other hand, through the use of standard numerical methods, we found the solutions of the model and we transpose techniques from the study of magnetic domains to estimate their size and shape as functions of  the material parameters mentioned above. 
\\

Finally, we would like compare the studies about the spherulites with those concerning the extended  solitonic configurations in chiral nematics. In particular, an analysis of the helicoidal configurations raising in confined CLCs, namely cholesteric fingers, can be carried out to obtain as equilibrium equation the elliptic Sine-Gordon on the strip \cite{key-1}. Very detailed studies of the solutions of this latter equation are known in the literature  \cite{Borisov,Novok},  in general on the whole plane or  with  boundary conditions significantly simpler then  the hometropic anchoring conditions, i.e. eq. \eqref{surfcond2}. However, it should be stressed that all nice solutions of the Sine-Gordon equation come out from its integrability properties,  studied  in \cite{Leib,Borisov3}, where the boundary conditions enter  in defining the Inverse Spectral Transform to a large extent. \\
A possible reduction compatible with the above requirements is   to look for solutions depending by separated $x$ and  $z$ variables.    
This idea was already applied several times  \cite{Lamb,Aero,Burylov, Zagro} and it can be implemented by  the  assumption
$  \theta = 4 \arctan \lq X\lf x \rg Z\lf z \rg \rq$. Boundary conditions \eqref{surfcond2} give rise to line disclinations on the confining planes, as analysed in \cite{17} in the absence of external fields. Thus, one should look for functions $X(x)$ and $Z(z)$ such that the former is monotonic and unbounded (with possible singularities at finite points) and the latter assume the value $Z\lf \pm \frac{L}{2} \rg = \textit{const}$ on the boundaries. On the other hand, more general solutions on the semi-strip with suitable \emph{integrable }boundary conditions have been built \cite{FLP}, often in a quite implicit way. Thus some additional work is needed in order to extract from them physical detailed informations. The aspects just described about the cholesteric fingers will be presented and discussed in a future paper.

\end{document}